\documentclass[%
reprint,
superscriptaddress,
noeprint,
 amsmath,amssymb,
 apl,
floatfix,
]{revtex4-2}
\expandafter\let\csname equation*\endcsname=\relax 
\expandafter\let\csname endequation*\endcsname=\relax

\usepackage{amsmath}

\usepackage[colorlinks = true,
            linkcolor = black,
            urlcolor  = blue,
            citecolor = black,
            anchorcolor = black]{hyperref}
            
\usepackage[modulo]{lineno}

\usepackage{graphicx}
\newlength{\figw}
\usepackage{hyperref}
\usepackage{verbatim}
\usepackage{siunitx}
\usepackage{xcolor}
\usepackage[utf8]{inputenc}

\DeclareSIUnit{\fluence}{\milli\joule\per\centi\meter\squared}

\DeclareSIUnit[]\muBohr
{\text{\ensuremath{\mu_{\textup{B}}}}}

\newcommand{\icol}[1]{
  \left(\begin{smallmatrix}#1\end{smallmatrix}\right)%
}

\newcommand{\matr}[1]{\mathbf{#1}} 

\newcommand{\iu}{\mathrm{i}\mkern1mu}

\begin{document}



\title[Wide-field magneto-optical microscope]{Wide-field magneto-optical microscope to access quantitative magnetization dynamics with femtosecond temporal and sub-micrometer spatial resolution}
\author{F.~Steinbach}
\author{D.~Schick}%
\author{C.~von~Korff~Schmising}\email{korff@mbi-berlin.de}
\author{K.~Yao}%
\author{M.~Borchert}%
\author{W.~D.~Engel}%
\affiliation{ 
Max-Born-Institut fuer Nichtlineare Optik und Kurzzeitspektroskopie, Max-Born-Strasse 2A, 12489 Berlin, Germany
}%
\author{S.~Eisebitt}%
\affiliation{ 
Max-Born-Institut fuer Nichtlineare Optik und Kurzzeitspektroskopie, Max-Born-Strasse 2A, 12489 Berlin, Germany
}%
\affiliation{ 
Institut fuer Optik und Atomare Physik, Technische Universitaet Berlin, Strasse des 17. Juni 135, 10623 Berlin, Germany
}%

\date{\today}


\begin{abstract}
We introduce a wide-field magneto-optical microscope to probe magnetization dynamics with femtosecond temporal and sub-micrometer spatial resolution.
We carefully calibrate the non-linear dependency between the magnetization of the sample and the detected light intensity by determining the absolute values of the magneto-optical polarization rotation. 
With that, an analytical transfer function is defined to directly map the recorded intensity to the corresponding magnetization, which results in significantly reduced acquisition times and relaxed computational requirements.
The performance of the instrument is characterized by probing the magnetic all-optical switching dynamics of GdFe in a pump-probe experiment. 
The high spatial resolution of the microscope allows for accurately subdividing the laser-excited area into different fluence-regions in order to capture the strongly non-linear magnetization dynamics as a function of the optical pump intensity in a single measurement.
\end{abstract}


\maketitle


\section{Introduction}


More than two decades after the seminal work of Beaurepaire et al. demonstrating the sub-picosecond quenching of magnetization after optical excitation \cite{Beaurepaire1996}, the field of ultrafast magnetism has remained a topic of intense research with frequent discoveries of new phenomena.
Laser pulse excitation has been shown to enable novel and energy-efficient pathways to manipulate magnetic order, ranging from ultrafast demagnetization, spin injection across interfaces \cite{Rudolf2012a,Melnikov2011,Kampfrath2013}, spatially controlled response in micropatterned media \cite{Savoini2012,VonKorffSchmising2014, LeGuyader2015},  magnetization reversal via all-optical switching (AOS) \cite{Stanciu2007,Vahaplar2009} to the controlled nucleation of magnetic skyrmions \cite{Finazzi2013,Buttner2021}.
The latter examples are processes that involve the transition to a state with an altered magnetic domain configuration, and, hence, require experimental techniques that give access to the lateral spatial dimension. 
Next to scanning techniques utilizing a tightly focused laser beam \cite{Freeman1996,Laraoui2007}, wide-field magneto-optical (MO) imaging based on the Kerr or Faraday Effect has become the most common approach to access spatial information with sub-micrometer to macroscopic length scales \cite{McCord2015}. Early applications, reaching nanosecond to picosecond temporal resolutions, studied magnetization dynamics in recording head writers \cite{Chumakov2005} and domain wall oscillations of a patterned ferromagnetic film \cite{Mozooni2014}. Recently,  wide-field MO microscopy has been increasingly used to investigate ultrafast magnetization dynamics and AOS \cite{Beens2019,Ceballos2021, Banerjee2020a} offering a time resolution only limited by the pulse duration of the femtosecond optical laser pulses~\cite{Vahaplar2009,Vahaplar2012,Savoini2012,Hashimoto_2014,Tsema2016,Stupakiewicz2017, Wang2021}. 
However, retrieving \textit{quantitative} information from MO images, a prerequisite to study ultrafast magnetization dynamics, has only been rarely reported in literature \cite{Laraoui2007,Hashimoto_2014}, mostly because it generally requires time-consuming polarization analysis and high performance computing. 
Finally, as optical excitation pulses typically exhibit a spatial intensity profile, a further advantage of MO imaging techniques is the possibility to accurately map the intensity dependence of laser-driven magnetization processes \cite{Hashimoto_2014}. 

In this work, we describe a time-resolved, wide-field microscope setup exploiting the MO Faraday or Kerr Effect.
The setup is optimized to probe the ultrafast magnetization dynamics of samples with perpendicular magnetic anisotropy obtaining femtosecond temporal and sub-micrometer spatial resolution.
By acquiring microscope images as a function of the angular orientation of an optical analyzer, we extract absolute values of spatially resolved polarization rotation angles, which are directly proportional to the sample's magnetization~\cite{Laraoui2007, Hashimoto_2014}.
By this, we derive a sample specific analytical transfer function to obtain an explicit expression for the dependence of the detected light intensity and the sample's magnetization. 
Such a transfer function has to be calculated for each sample material for which a different magneto-optical response can be expected. 
For time-resolved measurements this formalism renders the analyzer scans obsolete and effectively reduces the total acquisition time without the need for extensive computational requirements.
We exploit the Gaussian intensity distribution of the pump laser pulses to record a large range of excitation fluences simultaneously, revealing a highly non-linear relationship between the evolution of the magnetization and the excitation fluence.
We benchmark our experimental approach by probing the time- and fluence-dependent AOS in ferrimagnetic GdFe alloy.

\section{Experimental setup}

\begin{figure}[!ht]
    \centering
    \includegraphics[width=\linewidth]{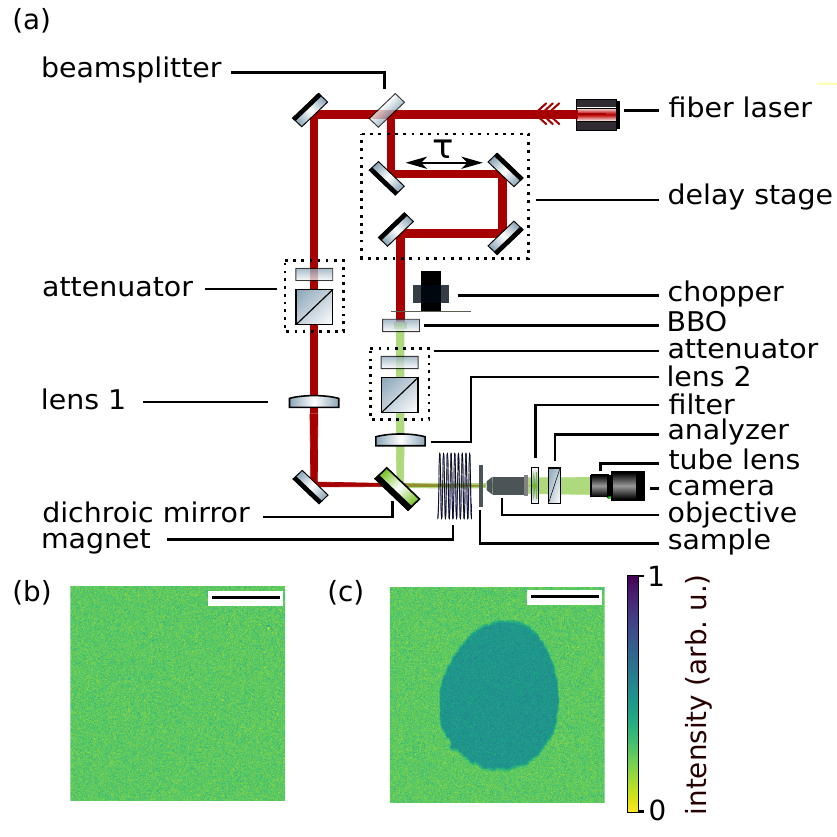}
    \caption{(a)~Wide-field magneto-optical microscope setup. 
    A femtosecond fiber laser is used as a light source for the microscope. 
    For time-resolved measurements, a pump/probe configuration is used.
    The pump pulses excite the sample at the fundamental laser wavelength of \SI{1030}{\nano\meter}, while the magnetic state of the sample is probed with the second harmonic at a wavelength of \SI{515}{\nano\meter}. 
    (b)~Intensity image of a saturated sample corresponding to the out-of-plane magnetization direction $M_+$. 
    (c)~Intensity image of the sample after excitation with a single pump pulse. 
    The dark color corresponds to a reversed magnetization direction, $M_-$. 
    Both images were recorded in thermal equilibrium long after optical excitation.
    The scale bar in both images corresponds to \SI{15}{\micro\meter}.}
    \label{fig:Fig1}
\end{figure}

An overview of the wide-field, time-resolved MO microscope in Faraday geometry is sketched in Fig.~\ref{fig:Fig1}~(a).
An Yb-based fiber laser~(Amplitude Laser Group, Satsuma HP) with a central wavelength of \SI{1030}{\nano\meter} and a variable pulse length between \SI{250}{\femto\second} to \SI{10}{\pico\second} is used as the light source. 
The laser pulses have a maximum pulse energy of \SI{20}{\micro\joule} at a variable repetition rate between single shot and \SI{500}{\kilo\hertz}, set by an acousto-optical modulator.
We implement a pump-probe geometry by splitting the laser pulses into two replicas with an intensity ratio of 30/70 (transmission/reflection). 
A mechanical delay stage (Newport, FMS300cc) controls the arrival time of the probe versus the pump pulses, allowing for delays between \SI{-10}{\pico\second} and \SI{1.7}{\nano\second} with a bi-directional repeatability of about \SI{10}{\femto\second}.
The beampath of the pump pulses is kept fixed in order to avoid any pointing instabilities, as their footprint on the sample is much smaller compared to the probe pulses.
The latter are guided via the mechanical delay stage and are actively stabilized to minimize spatial fluctuations of the illumination.
The probe pulses are frequency doubled to a wavelength of \SI{515}{\nano\meter} in a barium borate~(BBO) crystal and focused onto the sample to homogeneously illuminate the field of view of approximately \SI{200}{\micro\meter}. 
The pump pulses are focused by a separate lens onto the sample and a bandpass filter (Thorlabs, FESH0750) prevents them from entering the camera.
A dichroic mirror in front of the sample with high reflectivity for \SI{515}{\nano\meter} and high transmissivity for \SI{1030}{\nano\meter} ensures full collinearity of the pump and probe pulses.
Both optical paths are equipped with attenuators consisting of a motorized half-waveplate and a fixed Glan-Thompson polarizer with an extinction~$> 10^5$. 
The motor rotary stages (Newport, CONEX-AG-PR100P) allow for a minimum angular step size of \SI{1}{\milli\degree} with a bi-directional repeatability of \SI{3}{\milli\degree}.
Rotating the waveplate allows for accurately setting the pulse energy between \SIrange{0.002}{0.7}{\micro\joule} for the probe beam and between \SIrange{0.01}{4}{\micro\joule} for the pump beam. 
The attenuator is calibrated with a commercial powermeter (Thorlabs, S120C).

In Faraday geometry the polarization images of the samples are projected by an objective lens~(Zeiss EC~Epiplan-NEOFLUAR~50X, NA~0.8) and a tube lens onto a CMOS camera (Hamamatsu C13440-20CU). 
The dichroic, complex index of refraction of the magnetic samples causes changes of the polarization state of the probe light, both, in ellipticity and orientation.
The rotation of the plane of polarization is quantified by the so-called Faraday angle $\theta_\text{F}$ and is directly proportional to the magnetization of the sample.
An MO image emerges in a crossed-polarizer geometry, where a motorized analyzer optic (Thorlabs, LPVISA100) with an extinction of~$> 10^7$ behind the sample is set close to 90$^\circ$ with respect to the direction of the incoming light polarization.
The camera then only detects light with an altered polarization state due to transmission through magnetically ordered regions of the sample.

The magnetic state of the samples can be controlled by an external electromagnet with a maximum out-of-plane field strength of \SI{300}{\milli\tesla} at the sample position.
In addition to hysteresis loops, the external magnetic field is applied during pump-probe measurements to re-saturate the samples and to enable field-dependent studies.
The later point is of importance for deterministic AOS experiments, where every odd pump pulse initiates AOS and every even pump pulse resets the sample to its initial magnetic state without requiring an external magnetic field.
To record the transient magnetization dynamics a mechanical chopper is synchronized to block the even probe pulses, allowing to follow the dynamics induced by odd pump pulses only at half the available laser repetition rate ~\cite{Schick_2020}.

The setup can be easily adapted to change between a transmission (Faraday) and reflection (Kerr) geometry, e.g. for the investigation of optically opaque samples.
In reflection geometry, the pump pulse is focused by the imaging microscope objective, while the light path of the probe pulse is aligned according to a Köhler illumination leading to a uniform light distribution on the sample.
A continuous-flow cryostat is available to control the sample temperature between \SIrange{10}{450}{\kelvin}.
For temperature-dependent measurements, we account for the finite size of the evacuated cryostat by using a long-working distance objective~(Edmund Optics, 50X EO Long Working Distance) providing a spatial resolution of approximately \SI{1}{\micro\meter}.
We use the open-source software Sardana~\cite{sardana} to control the fully automatized setup, which can operate in a unsupervised mode by running through user-defined macro files.
This allows for systematic measurements, automatically scanning different parameters like temperature, fluence, or the external magnetic field.

We demonstrate the performance of the setup by investigating a magnetic Gd$_{24}$Fe$_{76}$(\SI{20}{\nano\meter}) alloy sandwiched between two Ta(\SI{5}{\nano\meter}) layers, grown onto a glass substrate by DC magnetron sputtering. 
Its magnetization is perpendicular to the sample plane with a coercive field of \SI{15}{\milli\tesla}.
The sample exhibits AOS which is evidenced by the two magneto-optical images shown in Fig.~\ref{fig:Fig1}~(b) and (c), which are recorded before and after single pulse excitation in thermal equilibrium.
The magnetic contrast for the saturation magnetization~$M_+$ is given by green and the area with reversed magnetization~$M_-$ due to AOS is encoded by a blue color. 
We confirm a spatial resolution of below \SI{1}{\micro\meter}, by evaluating the lineout between the two oppositely magnetized domains and defining the resolution as the width between 90\% and 10\% of the total intensity contrast.

\subsection{Faraday Rotation Analysis}

We apply the Jones calculus to establish the relationship between the detected light intensity on the camera and the Faraday angle,~$\theta_\text{F}$. 
The probe pulse is described as the electric field vector $\Vec{E}_\text{in}=\icol{E_x\\E_y}$, where $E_x$ and $E_y$ are the complex horizontal ($x$) and vertical ($y$) components of the field. 
The polarization-dependent optical elements are described by matrices and consist of the horizontally aligned polarizer~$\matr{P}$ and the magnetic sample~$\matr{S}$ as well as components that contribute to parasitic Faraday rotations~$\matr{F}$, and finally the analyzer~$\matr{A}$, aligned at an angle~$\alpha$ with respect to incoming polarization direction, c.f. Eq.~\ref{eqn:jones}.
After transmission through the sample, the probe light polarization exhibits a magnetization-dependent Faraday rotation, $\theta_\text{F}$, and change of ellipticity, $\eta$.
The parasitic Faraday angle, $\phi$, depends on the external magnetic fields, $B_\text{ext}$, and is mainly caused by the glass lenses inside the objective \cite{Ishibashi2004}.

\begin{eqnarray}
 \matr{P}
 & = &
  \begin{bmatrix}
   1 & 0 \\
   0 &0 
   \end{bmatrix} \nonumber\\
 \matr{S} 
 & = &
  \begin{bmatrix}
   \cos{\theta_\text{F}}+\iu\eta\sin{\theta_\text{F}} &  -\sin{\theta_\text{F}}+\iu\eta\cos{\theta_\text{F}}\\
    \sin{\theta_\text{F}}-\iu\eta\cos{\theta_\text{F}} & \cos{\theta_\text{F}}+\iu\eta\sin{\theta_\text{F}} 
   \end{bmatrix} \nonumber\\   
 \matr{F}
 & = &
  \begin{bmatrix}
   \cos{\phi} &  -\sin{\phi}\\
    \sin{\phi} & \cos{\phi} 
   \end{bmatrix}
 \nonumber\\
 \matr{A}
 & = &
  \begin{bmatrix}
   \sin^2{\alpha} & \sin{\alpha}\cos{\alpha} \\
   \sin{\alpha}\cos{\alpha} & \cos^2{\alpha} 
   \end{bmatrix}
   \label{eqn:jones}
\end{eqnarray}
In front of the camera the electric field vector is given by $\Vec{E}_\text{out}=\matr{A}\matr{F}\matr{S}\matr{P}\Vec{E}_\text{in}$.
The intensity, $I$, detected by the camera is then calculated by \cite{Hashimoto_2014}
\begin{eqnarray}
I =|\Vec{E}_\text{out}|^2 & = & I_0\left[\left(1-\eta^2\right) \sin^2{\left(\alpha+\theta_\text{F}+\phi\right)}+\eta^2\right] \nonumber\\
  & \approx & I_0'\left(\alpha+\theta_\text{F}+\phi\right)^2+I_\mathrm{\epsilon}
\label{equ:intens}
\end{eqnarray}
where $I_0$ is the incoming intensity and $I_\mathrm{\epsilon}$ is an ellipticity depended offset. 
The final step of Eq.~\ref{equ:intens} corresponds to a small-angle approximation. 
Evidently, the detected intensity depends non-linearly on the Faraday angle and therefore on the magnetization of the sample. 

\begin{figure}[!ht]
    \centering
    \includegraphics[width=\linewidth]{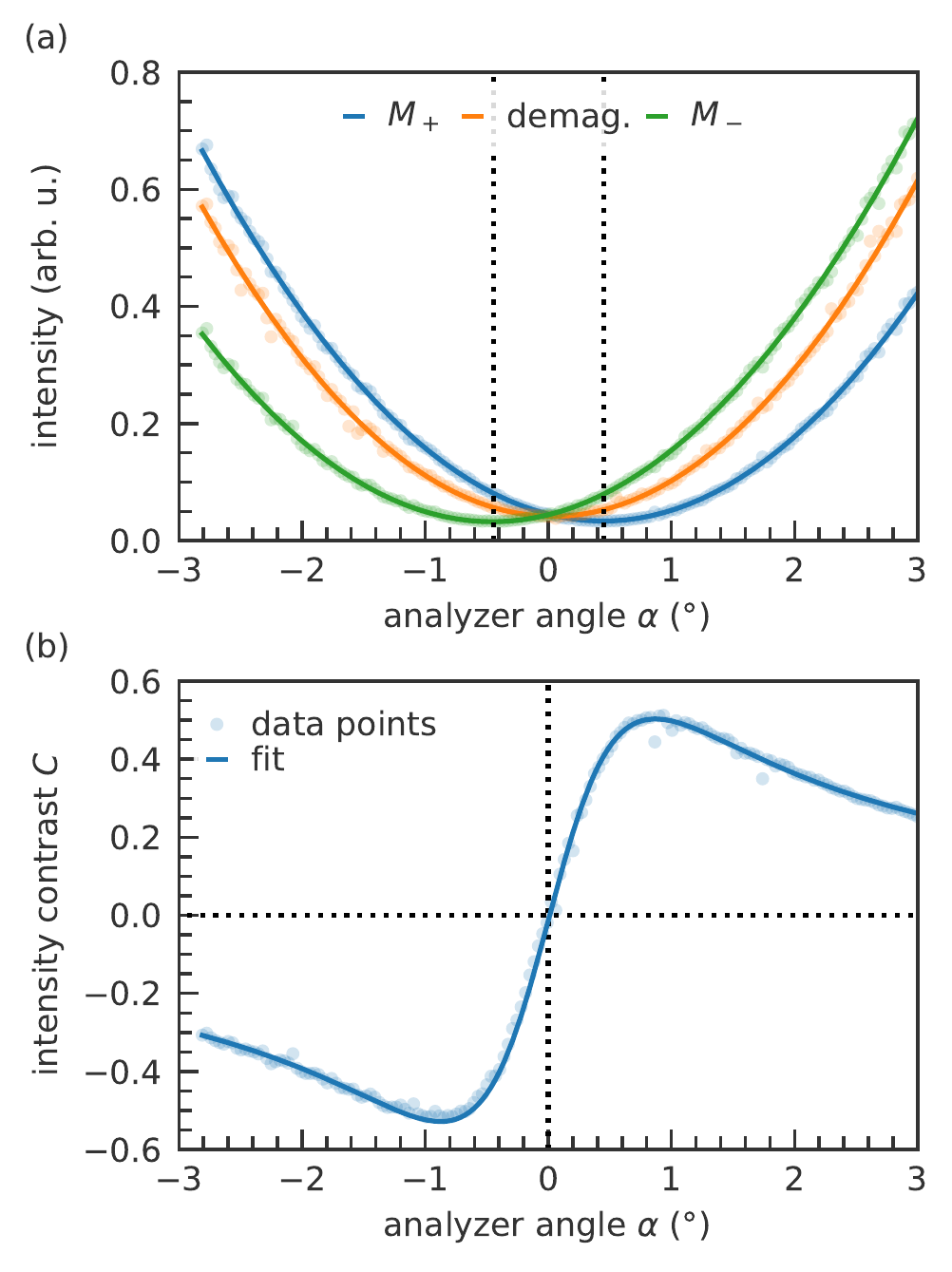}
    \caption{(a) Scan of the analyzer angle, $\alpha$, for three different magnetic states, $M_+$ saturation (blue), $M_-$ saturation (green) and  $M_0$ (orange), i.e. a fully demagnetized state where the  sample’s magnetization projection averaging to  zero in the out-of-plane direction. 
    The shift of the parabolic fit functions is caused by the Faraday rotation of the different magnetic states of the sample. The vertical dotted lines mark the Faraday rotation of the sample, $\pm\theta_\text{sample} = \pm 0.45^\circ$
    (b) Intensity contrast for different analyzer angles with maxima at $\alpha = \pm 0.9^\circ$. The contrast is calculated according to Eq.~\ref{equ:contrast} using the data points and the fits shown in panel (a).}
    \label{fig:fig2}
\end{figure}

We determine the absolute Faraday rotation of the sample by scanning the analyzer angle, $\alpha$, for opposite out-of-plane magnetization directions, $M_\pm$, as well as for a demagnetized state $M=0$, i.e. for an average over a magnetic multi-domain pattern, as shown in Fig.~\ref{fig:fig2}~(a).
We fit the data with the quadratic function according to Eq.~\ref{equ:intens}.
The different positions of the minima of the parabolas for measurements with opposite magnetization directions, $M_\pm$, are marked with dotted vertical lines and determine the maximum, sample specific Faraday rotation, $\pm\theta_\text{sample} = \pm 0.45^\circ$.
The Faraday rotation of the demagnetized state (orange line) is zero and located between the response of $M_-$ and $M_+$ (blue and green lines, respectively). 
The intensity offset is caused by the ellipticity, $\eta$, of the sample, which also depends on the magnetization. 
Importantly, in this plot parasitic Faraday rotations, $\phi$, only lead to a constant offset on the $\alpha$ axis, independent of the magnetic state of the sample and do not influence $\theta_\text{sample}$. Subtraction of the parasitic Faraday rotation ensures that we find the analyzer angle $\alpha = 0$ as the minimum of Eq.~\ref{equ:intens} for the demagnetized state.
These measurements allow us to determine the intensity contrast according to
\begin{gather}
C(\alpha)=\frac{I(M_+,\alpha) - I(M_-,\alpha)}{I(M_+,\alpha) + I(M_-,\alpha)}
\label{equ:contrast}
\end{gather}
For the GdFe alloy sample the maximum contrast is at $\pm\, 0.9^\circ$  and is determined by the values of $\theta_\text{sample}$, the incoming intensity $I_0$, and the offset $I_\mathrm{\epsilon}$, see Fig.~\ref{fig:fig2}~(b). We note that the highest signal to noise ratios are achieved for the analyzer angles which maximize $C$. 

\subsection{Magnetization Normalization}

To normalize the Faraday rotation to a corresponding magnetization two additional analyzer scans are needed. 
One scan is performed with a saturated sample for one magnetization direction, $M_{+/-}$, set by an external magnetic field, $B_{+/-}$. 
Another analyzer scan is performed with a blank substrate applying the same external magnetic field as used in the previous measurement, $B_{+/-}$, to extract the parasitic Faraday rotation~$\phi$. 
This yields the absolute value of $\theta_\text{sample}$. 
Note, that for measurements without external magnetic field only the saturated $\alpha$-scan is necessary as $\phi$ equals zero.

The Faraday angles, $\pm \theta_\text{sample}$, correspond to the saturation magnetizations $M_\pm$ such that we can calculate the transient magnetization normalized between -1 and 1 according to
\begin{gather}
    M(t) = \frac{\theta_\text{F}(t)+\theta_\text{sample}}{\theta_\text{sample}}-1
    \label{equ:calib}
\end{gather}

Therefore, the measurements not only yield absolute values of the Faraday rotation, but also allow to follow relative changes of the magnetization.

\section{All-optical Magnetic Switching}

\begin{figure}[!ht]
    \centering
    \includegraphics[width=\linewidth]{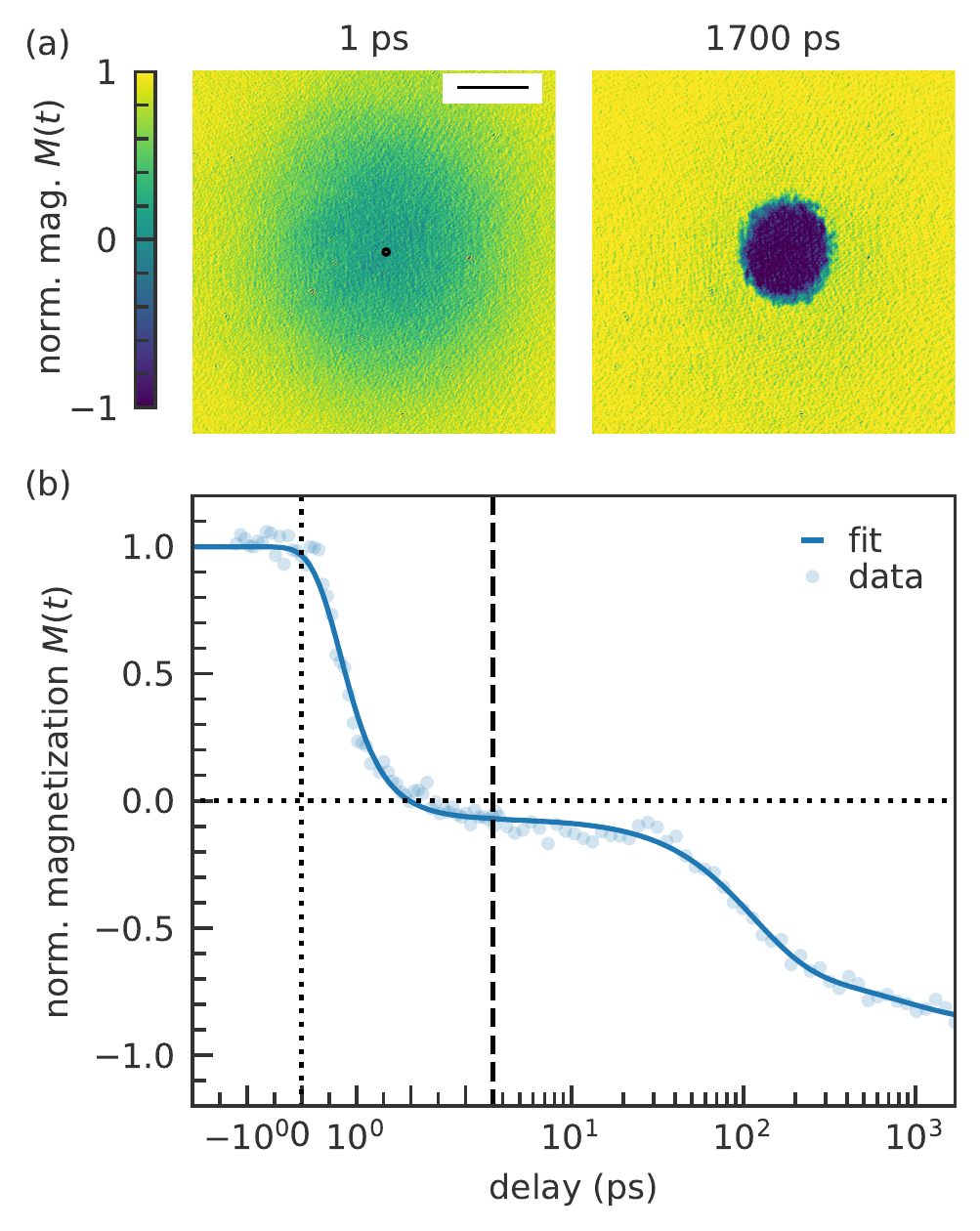}
    \caption{(a) Magnetization of the GdFe sample at different delays after optical excitation. 
    The small black circle in first image (\SI{1}{\pico\second}) marks the integration area with a size of about \SI{0.2}{\micro\meter\squared} for the time-resolved transient shown in panel (b).
    The scale bar corresponds to \SI{25}{\micro\meter}.(b) Transient magnetization of the sample after laser excitation for a fluence of $\SI{4.9}{\milli\joule\per\centi\meter\squared}$ extracted by integrating the central area shown in the left panel of (a). Note that the time axis is linear between \SI{-2}{\pico\second} and \SI{3.5}{\pico\second} and logarithmic for later times up to \SI{1.7}{\nano\second}. The transition is marked by the dashed vertical line.
    } 
    \label{fig:fig3}
\end{figure}

We perform time-resolved experiments to measure the magnetization dynamics of the GdFe alloy after optical excitation. To this end, we record analyzer scans for each time delay and extract the associated Faraday rotations.
In Fig.~\ref{fig:fig3} (a), we show two MO images for selected time delays at $t= $\SI{1}{\pico\second} and $t= $\SI{1700}{\pico\second}. 
The background of the images are corrected by subtraction of an image recorded before time delay zero at $t=\SI{-2}{\pico\second}$. 
At a time delay of \SI{1}{\pico\second} a circular area with a diameter of approximately \SI{30}{\micro\meter} is fully demagnetized. The edges of the field of view correspond to very weakly excited areas and, hence, have almost remained in the initial magnetization state, $M_+$. 
After \SI{1700}{\pico\second} only a small switched spot ($M_-)$ with a diameter of approximately \SI{15}{\micro\meter} remains, while the external magnetic field has almost completely reset the magnetization of the surrounding areas.

Integration over a circular area positioned at the center of the excitation with a diameter of \SI{0.5}{\micro\meter} (black circle in Fig.~\ref{fig:fig3} (a)) yields the magnetization dynamics for the maximal excitation fluence $F_\text{max}= (4\ln{2}\cdot \varepsilon)/(\pi \cdot \delta^2) =  \SI{4.9}{\milli\joule\per\centi\meter\squared}$, where $\varepsilon$ is the pulse energy and $\delta$ corresponds to FWHM of the laser spot. The corresponding values of the magnetization as a function of time are shown in Fig.~\ref{fig:fig3}~(b). 
For this measurement, we recorded MO images for 12 distinct analyzer angles, $\alpha$, with an exposure time of \SI{100}{\milli\second} at each of the 113 time delay points. Taking into account additional time required to move the different motors and read out the CCD images, this measurement took approximately 16~minutes. 
The magnetization dynamics is fitted with a bi-exponential function and convolved by a Gaussian function to account for the time resolution of our experiment \cite{radu2015}.
The time resolution is determined independently by a cross correlation of the pump and probe pulses, which for pulse durations of \SI{250}{\femto\second}, amounts to approximately \SI{350}{\femto\second}.\\
The magnetization of the GdFe alloy decreases very quickly after excitation and reverses its sign in less than two picoseconds. 
Then the magnetization reaches a switched state of $0.8\cdot M_-$ after approximately \SI{400}{\pico\second}.

\begin{figure}[!ht]
    \centering
    \includegraphics[width=\linewidth]{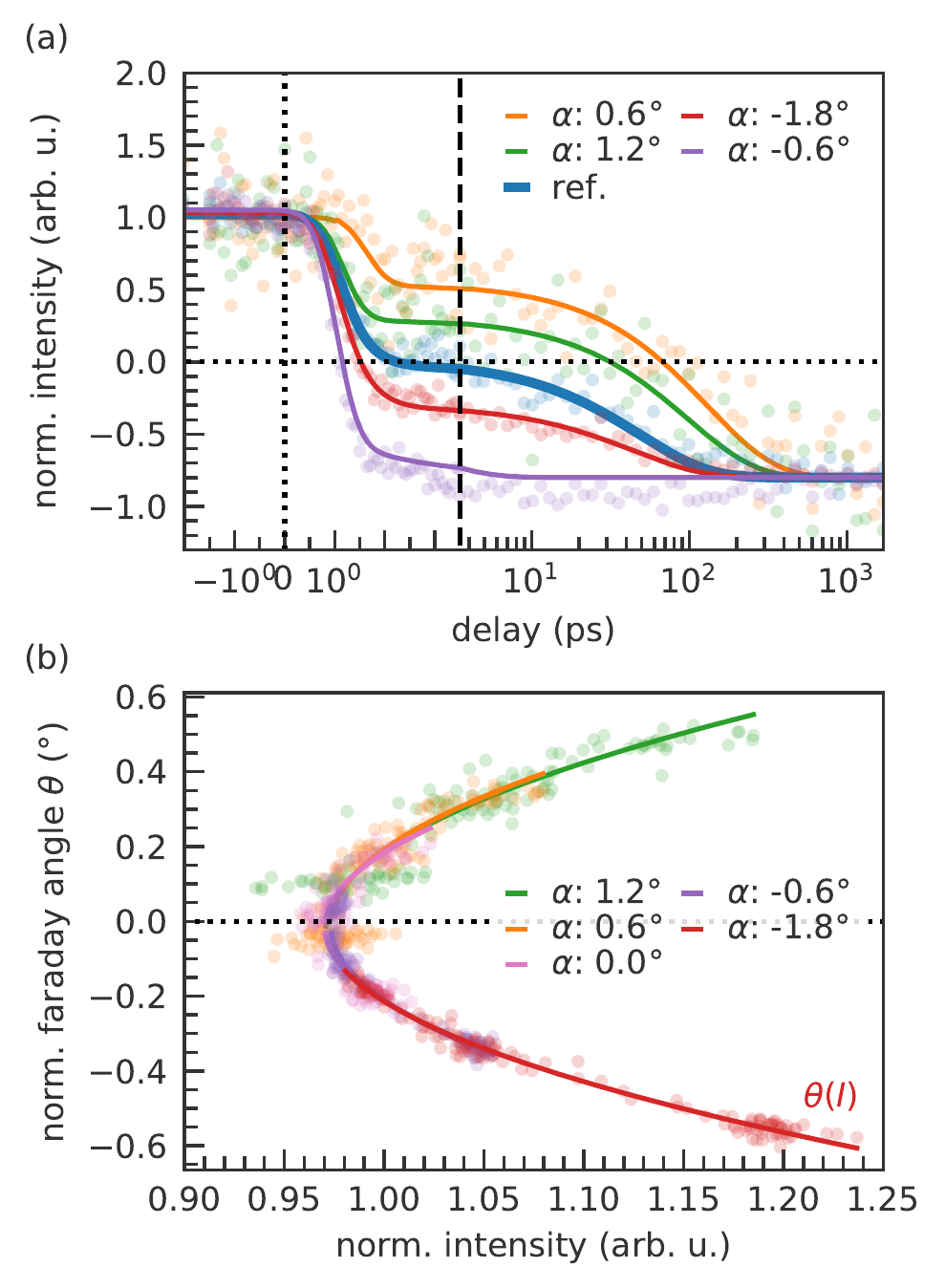}
    \caption{(a) Normalized intensities as a function of time delay measured for different fixed analyzer angles $\alpha$. 
    The very different response highlights the non-linear relation between intensity and Faraday rotation.
    (b) Normalized Faraday angle as a function of the time-dependent intensity. For a sufficiently large Faraday angle e.g. $\alpha$ = \SI{-1.8}{\degree}, a unique transfer function $\theta(I)=0.8(I+0.02)^2+1$ can be defined (red line). 
    } 
    \label{fig:fig4}
\end{figure}

To further demonstrate the importance to measure normalized Faraday angles in order to extract quantitative magnetization dynamics from MO images, we plot transients for different fixed analyzer angles in Fig.~\ref{fig:fig4}~(a). The region of interest is identical to our previous analysis and corresponds to  the circular area with a diameter of \SI{0.5}{\micro\meter} as shown in the left MO image of Fig.~\ref{fig:fig3}~(a). 
We normalize the intensities by assuming that the intensity before time delay zero is equal to one and after a delay of \SI{400}{\pico\second} equal to $-0.8$, in accordance with the magnetization dynamics shown in Fig.~\ref{fig:fig3}~(b).
This assumption would be correct if quantitative magnetization dynamics could be extracted for arbitrary analyzer angles provided they lead to a finite intensity contrast. 
As a reference, we additionally plot the transient \textit{magnetization} as shown in Fig.~\ref{fig:fig3}~(b) as a solid blue line. 
Obviously, the intensity transients for different analyzer angles are very different, illustrating the strongly nonlinear relationship between detected intensity and magnetization. For example, for the negative analyzer angle, $\alpha= -0.6^\circ$, the normalized intensity values almost completely reverse their sign within \SI{2}{\pico\second}, while for the positive angle, $\alpha= 0.6^\circ$, the detected intensity crosses zero only after \SI{100}{\pico\second}. Interpreting the dynamical pathway of the magnetization between the two stable magnetic states, $M_\pm$, is clearly not possible without prior polarization analysis of the sample. This is an important message and should be considered when referring to published work on transient Faraday or Kerr microscopy \cite{Elazar2008,Vahaplar2009, Vahaplar2012,Savoini2012,Tsema2016,Stupakiewicz2017,Wang2021}.   
Furthermore, not only the apparent dynamics are different, but also the noise. 
This is expected when recalling how the intensity contrast, $C$, varies for different angles $\alpha$ as shown in Fig.~\ref{fig:fig2}~(b). 
In Fig.~\ref{fig:fig4}~(b), we plot the Faraday angle for five different values of the analyzer angle, $\alpha$, as a function of the normalized time-dependent intensities after optical excitation. For small analyzer angles the Faraday angle passes through the minimum as the sample is reversing its magnetization and the measured intensities do not allow to differentiate between $M_+$ and $M_-$. However, for a sufficiently large analyzer angle, $\alpha = \SI{-1.8}{\degree}$, every intensity can be assigned to a unique Faraday angle. 
This allows the definition of a transfer function $\theta(I)$ (red line in Fig.~\ref{fig:fig4} b) and establishes an analytical relationship between measured intensity and Faraday angle. 
Note that setting the analyzer angle to $\alpha \approx \pm\SI{0.6}{\degree}$, i.e. close to the maximum contrast, $C$ (cf. Fig.~\ref{fig:fig2}~(b)), no unique transfer function can be defined.
We like to point out that the sample specific Faraday-intensity relationship can not only be retrieved in time-resolved measurements, but in principle in any experiment where the variation of an external parameter allows to continuously change the magnetization of a sample, i.e. by temperature-dependent measurements.\\ 
The determination of a unique transfer function implies that further measurements for this sample do not require analyzer scans at each time delay to extract the correct magnetization dynamics. 

\section{Fluence Dependence}

\begin{figure}[!ht]
    \centering
    \includegraphics[width=\linewidth]{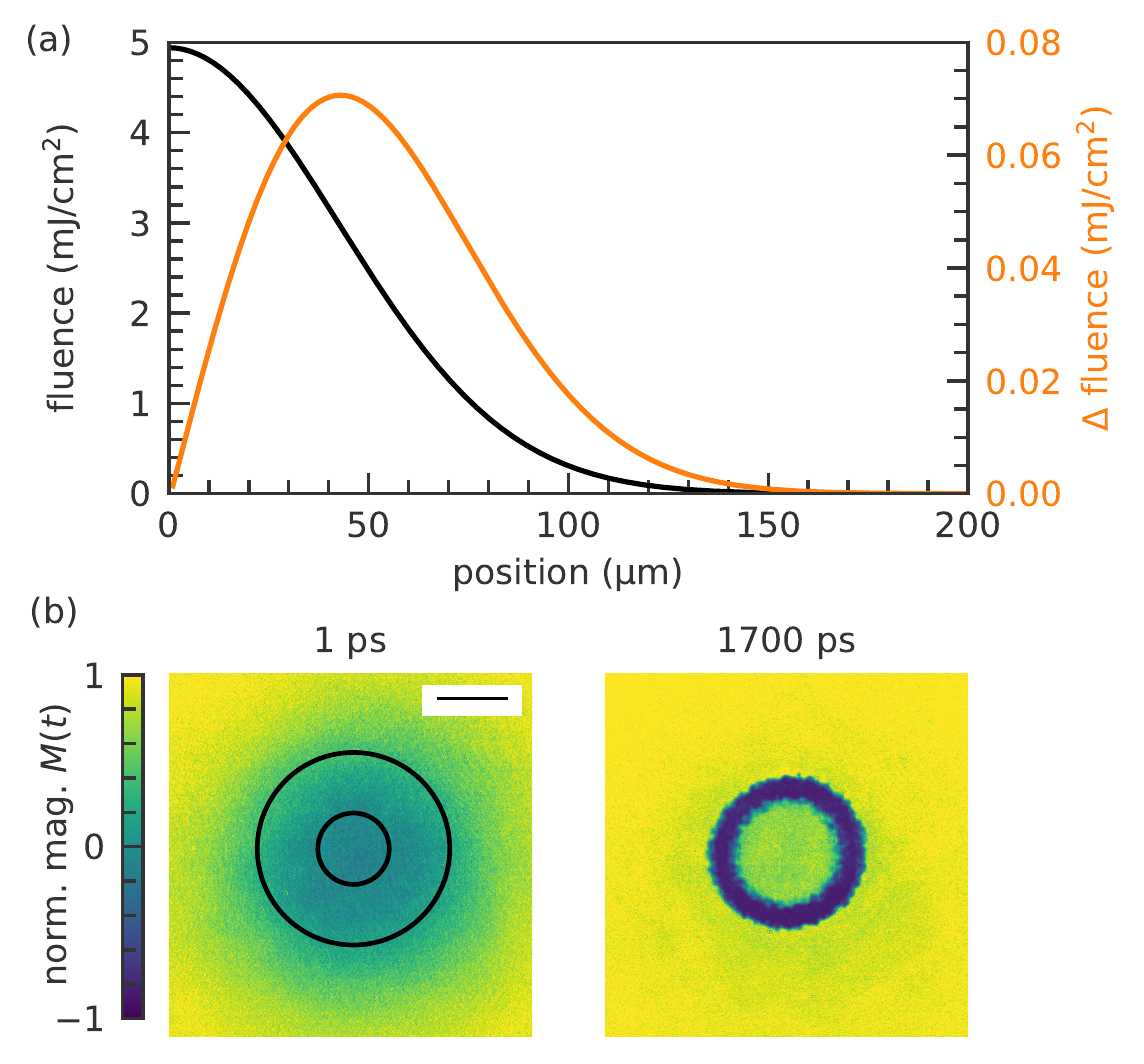}
    \caption{(a) Variation of the fluence (black line) and of the fluence resolution (orange line) as a function of spatial position, $R$. (b) Normalized magnetization maps for different delays after excitation with a peak fluence of \SI{4.9}{\fluence}. Pixels on the circular segments (black lines shown in image for $t = 1$~ps) correspond to positions that are excited by the same fluence. The inner ring corresponds to a fluence of \SI{4.7}{\fluence}, the outer ring to \SI{3.6}{\fluence}. The scale bar corresponds to \SI{25}{\micro\meter}.} 
    \label{fig:fig5}
\end{figure}

The possibility to locally analyse the magnetic state within microscope images allows us to measure the fluence dependence of the magnetization dynamics within a single acquisition.
As the spatial intensity distribution of our laser pulse exhibits a Gaussian shape with a FWHM $\delta$, the fluence $F(R)$ with radius $R$ from the center of the beam varies according to  $F(R)=F_\text{max}\exp{(-8\ln{2}\cdot R^2/\delta^2)}$. 
For the following experiment, the pump pulse was focused to \SI{100}{\micro\meter} $\times$ \SI{100}{\micro\meter} FWHM, while the sample was probed with a beam of \SI{330}{\micro\meter} $\times$ \SI{400}{\micro\meter} FWHM. The FWHM were accurately determined with a beam profiler (Dataray, WinCamD-THz). For a maximum fluence $F_\text{max}$ of \SI{4.9}{\milli\joule\per\centi\meter\squared}, we show the variation of the fluence as a function of spatial position, $R$, in Fig.~\ref{fig:fig5} (a). 
In the same figure, we additionally plot the corresponding upper bound with which we can resolve different excitation fluences experimentally (orange line). Assuming a spatial resolution of $<$\SI{1}{\micro\meter}, we determine a maximum value of $<$\SI{0.07}{\milli\joule\per\centi\meter\squared}, spatially coinciding with the largest slope of the fluence distribution. We demonstrate the strongly nonlinear fluence dependence of AOS, by further investigating the magnetization dynamics of the GdFe sample.  In Fig.~\ref{fig:fig5} (b), we show the MO images calculated via the previously determined transfer function, $\theta(I)$, for two selected time delays. In the middle of the excitation distribution the fluence is the highest and decreases with increasing radius, $R$. We perform an azimuthal integration over pixels with the same fluence using the Phython package PyFAI, correcting for small deviations from a perfect radially symmetric excitation profile \cite{Kieffer2013}. Two exemplary integration pathways corresponding to \SI{4.7}{\fluence} and \SI{3.6}{\fluence} are drawn in the left panel of Fig.~\ref{fig:fig5} (b). 
We like to point out, that for inhomogeneous pump beam profiles or for excitation patterns that exhibit a tailored structure, the intensity distribution can also be directly recorded by the CCD camera of our imaging system. This allows to assign every pixel of the MO images a measured value of the pump intensity. 

\begin{figure}[!ht]
    \centering
    \includegraphics[width=\linewidth]{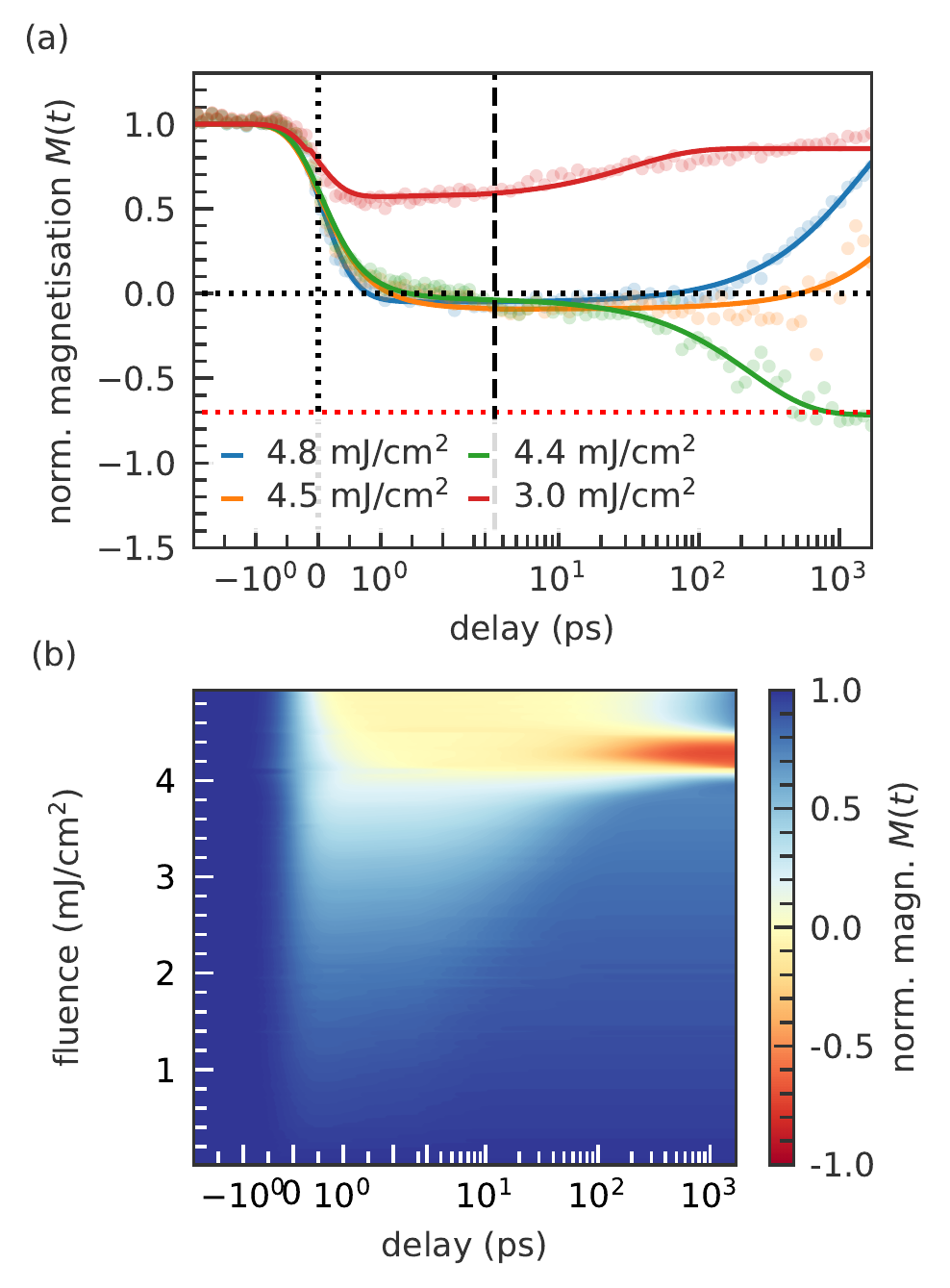}
    \caption{(a) Magnetic transients for different fluences which were extracted for different radii of the recorded images. The red dotted line marks a switching of \SI{70}{\percent}. (b) Fluence-delay-magnetization map for Gd$_{24}$Fe$_{76}$ sample. The map is a result of bi-exponential fits. The 2D map very clearly depicts the strongly nonlinear fluences dependence of AOS.} 
    \label{fig:fig6}
\end{figure}

In Fig.~\ref{fig:fig6}~(a) the transients for different fluences are shown. It is noteworthy that very small differences in fluence lead to significant quantitative and even qualitative changes of the magnetization dynamics.
Comparing the dynamics at a radius of \SI{11}{\micro\meter}~(\SI{4.8}{\fluence}) with dynamics at \SI{21}{\micro\meter}~(\SI{4.4}{\fluence}) shows a tremendously different dynamics although the difference of the two radii is only about \SI{10}{\percent} of the FWHM of the pump pulse.
This is an important observation, as for the more common approach of magneto-optical Kerr/Faraday Effect pump-probe experiments, using focused laser beams, the response within the probe spot is integrated, potentially bluring the differences between a fluence-dependent response.
For integrating MOKE experiments, our measurements show that it is crucial to ensure a sufficiently large ratio of the FWHM of pump and probe pulses in order avoid integration over distinct dynamics. 
For the investigated GdFe sample, the ratio should remain well below 1/10. 
To present the complete information contained in the series of time resolved MO images, we plot a 2D map of the fitted transient magnetization both as a function of fluence and time delay in Fig.~\ref{fig:fig6}~(b).
Three different fluence regions can be differentiated. Below \SI{4.2}{\fluence} the sample only demagnetizes transiently and relaxes back to its original magnetization direction.
Between \SI{4.2}{\fluence} and \SI{4.5}{\fluence} the sample completely switches its magnetization, ending up in a magnetically reversed state. 
For a fluence in excess of this value, the sample completely demagnetizes, but then resaturates to its initial state. Evidently, only for a relatively small fluence range the sample switches its magnetization completely.\\ 
The ability to generate such a fluence map shows the strength of a spatially resolving microscopic approach compared to a common used, spatially integrating MOKE/Faraday setup. As the entire fluence range is encoded in a single MO image, the fluence dependence on ultrafast magnetization dynamics can be carried out with high fidelity. With a known transfer function, $\theta(I)$, the fluence- and time-dependent magnetization map shown in Fig~\ref{fig:fig6}~(b) required an acquisition time of only 3~minutes.

\section{Conclusions}

We demonstrated a wide-field magneto-optical microscope setup to measure the magnetization via Faraday rotation with a femtosecond temporal and sub~\SI{1}{\micro\meter} spatial resolution. 
By performing analyzer angle scans for each time delay,  it is possible to extract the absolute value of the transient Faraday rotation for different regions of interests in the MO images. For an adequately chosen analyzer angle, we demonstrate that one can define a sample specific, analytic transfer function to calculate Faraday rotations from measured intensities, greatly reducing experimental acquisition times and computational requirements. We emphasize that a polarization analysis is a necessary requirement to extract quantitative statements about demagnetization and all-optical switching times from MO images.  Furthermore, we underlined the advantages of wide-field polarization microscopy allowing simultaneous measurements of different excitation fluences with a resolution of $<$\SI{0.07}{\fluence}. In future, we foresee further applications of our setup for the investigation of magnetization dynamics in chemically and magnetically inhomogeneous samples, for micropatterned magnetic systems or for studies employing a structured illumination.

\section{Acknowledgement}

We gratefully acknowledge financial support by the DFG through TRR227, project A02.

\section{Data Availability}
The data that support the findings of this study are openly available in Zenodo at http://doi.org/10.5281/zenodo.5152989.
\section{References}

\bibliography{references}

\end{document}